\documentclass[journal]{rmaa}

\newcommand{\1}{\'{\i}}

\begin{document}

\title{A new search for variable stars in the globular cluster NGC 6366\altaffilmark{1}}

\author{
  A. Arellano Ferro,\altaffilmark{2} 
  Sunetra Giridhar,\altaffilmark{3}
  V. Rojas L\'opez,\altaffilmark{2}
  R. Figuera,\altaffilmark{4}
  D.M. Bramich,\altaffilmark{5}
  and P. Rosenzweig.\altaffilmark{4}}

\altaffiltext{1}{Based on observations collected at the Indian
   Astrophysical Observatory, Hanle, India.} 
\altaffiltext{2}{Instituto de Astronom\1a, Universidad Nacional Aut\'onoma de M\'exico, M\'exico.}

\altaffiltext{3}{Indian Institute of Astrophysics, Bangalore, India.}

\altaffiltext{4}{Grupo de Astrof\1sica Te\'orica, Facultad de Ciencias, Universidad de Los Andes, M\'erida, Venezuela.}

\altaffiltext{5}{Isaac Newton Group of Telescopes, Santa Cruz de la Palma, Canary Islands, Spain.}

\fulladdresses{
\item A. Arellano Ferro, V. Rojas L\'opez: Instituto de Astronom\1a, Universidad Nacional Aut\'onoma de M\'exico, Apdo. Postal 70-264, M\'exico D. F. CP 04510, 
M\'exico. (armando@astroscu.unam.mx, victoria@astroscu.unam.mx)
\item D.M. Bramich: Isaac Newton Group of Telescopes, Apartado de Correos 321, E-38700 Santa Cruz de la Palma, Canary Islands, Spain. (dmb@ing.iac.es) 
\item Sunetra Giridhar: Indian Institute of Astrophysics, Bangalore, India.
\item R. Figuera, Patricia Rosenzweig: Grupo de Astrof\1sica Te\'orica, Facultad de Ciencias, Universidad de Los Andes, M\'erida, Venezuela. (rfiguera@ula.ve, patricia@ula.ve)
}

\shortauthor{Arellano Ferro, Giridhar, Bramich, et al.}
\shorttitle{Variable stars in NGC 6366}

\SetYear{2008}
\ReceivedDate{March 2008}
\AcceptedDate{------} 

\resumen{A trav\'es de fotometr\1a CCD de NGC 6366 se han descubierto nuevas estrellas variables. Se descubrieron: dos posibles Cefeidas An\'omalas (o Cefeidas de Pob II), tres gigantes de largo per\'iodo, una SX Phe y una binaria eclipsante. Tambi\'en se reporta una lista de 10 posibles variables. La curva de luz de la estrella RRab, V1, fue descompuesta en sus arm\'onicos de Fourier y \'estos se emplearon para estimar la metalicidad y la distancia de la estrella; [Fe/H]~$= -0.87 \pm 0.14$ y $d~=~3.2 \pm 0.1$ kpc. Se argumenta que la estrella V1 podr\1a no ser miembro del c\'umulo sino un objeto m\'as distante. Si es as\1, se puede emplear V1 para calcular un l\1mite superior para la distancia del c\'umlo de $2.8 \pm 0.1$ kpc.
La relaci\'on
$P-L$ para estrellas tipo SX Phe y los modos de pulsaci\'on identificados en la SX Phe descubierta, V6, permiten una determinaci\'on independiente de la distancia $d~=~2.7 \pm 0.1$ kpc.
El caso de V1 es discutido en el marco de la relaci\'on $M_V$- {\rm [Fe/H]} para estrellas RR Lyrae.}

\abstract{New CCD photometry of NGC~6366 has lead to the discovery of some variable stars.
Two possible Anomalous Cepheids (or Pop II Cepheids), three long period variables,
one SX Phe and one eclipsing binary have been found. Also
a list of 10 candidate variables is reported. The light curve of the RRab
star, V1, has been decomposed into its Fourier harmonics, and the
Fourier parameters were used to estimate
the star's metallicity and distance; [Fe/H]~$= -0.87 \pm 0.14$ and
$d~=~3.2 \pm 0.1$ kpc. 
It is argued that V1 may not be a member of the cluster but rather a more distant object.
If this is so, an upper limit for the distance to the cluster of $2.8 \pm 0.1$ kpc can be estimated.
The $P-L$ relationship for SX Phe stars and the identified modes in the
newly discovered SX Phe variable, V6,
allow yet another  independent determination of the distance to the
cluster of $d~=~2.7 \pm 0.1$ kpc.
The $M_V$- {\rm [Fe/H]} relationship for RR Lyrae stars is addressed and the case of V1 
is discussed.} 

\keywords{Globular Clusters-NGC 6366, Variable Stars - RR Lyrae, SX Phe}

\maketitle

\section{Introduction}
\label{sec:intro}

The importance of studying globular clusters is linked to the fact that they can provide insight into 
 the structure and evolution of the Galaxy. Hence, the determination of their ages, chemical compositions, galactocentric distances and kinematics are of fundamental relevance.  Numerous works in 
 recent literature  have been devoted to the estimation of the above mentioned quantities. 
Calculations of relative ages at 
a fixed metallicity provide an age scale with an uncertainty of $\leq$ 1 Gyr. This has allowed Harris et al. (1997) to estimate that the very metal-poor ([Fe/H] $\sim -2.14$) distant ($\sim 90$ kpc) cluster NGC 2419 has similar age to the inner cluster M92, and conclude that the earliest stellar globular cluster formation began at about the same time everywhere in the Galaxy. By the same technique Stetson et al. (1999) found that some of the outer-halo clusters ($R_{GC} \ge$ 50 kpc) (Palomar 3, Palomar 4 and Eridanus) are  
$\sim$1.5-2 Gyr younger
 than the inner-halo clusters, ($R_{GC} \leq$ 10 kpc) of similar metallicity
(M3 and M5). These author recognise however that the age difference can be
less than $\sim$ 1 Gyr if their [Fe/H] or [$\alpha$/H] abundances are overestimated. Similarly Sarajedini (1997) concluded that the outer-halo cluster Palomar 14 is 3 to 4 Gyr younger than inner clusters of similar metallicity.
It seems that most of the outer-halo globular clusters are 1.5-4 Gyr younger than the inner-halo clusters of similar [Fe/H], and that NGC 2419 may be an exception (VandenBerg 2000).

Based on their metallicities and kinematics, two populations of globular clusters have  been distinguished;
 a halo population composed of metal poor clusters (i.e. [Fe/H] $\leq -0.8$) which is slowly rotating ($V_{rot} = 20 \pm 29 $ km/s with a large line-of-sight velocity dispersion of $127 \pm 11$ km/s; Zinn 1996),
 and a disk
population composed of metal rich globular clusters (i.e. [Fe/H] $\ge -0.8$) which is rapidly rotating  ($V_{rot} = 157 \pm 26 $ km/s with a small line-of-sight velocity dispersion of $66 \pm 13$ km/s; Zinn 1996)  (see also Zinn 1985, Armandroff 1989; Zinn 1993; van den Bergh 1993). The metal-rich component may also include a bulge system ($R_{GC} \leq$ 3 kpc) which according to Minniti (1995) formed after the halo.
However, comparing the luminosity of the HB and the MS turn-off of the galactic bulge, the two metal-rich bulge globular clusters NGC~6528 and NGC~6553, and the inner Halo cluster NGC~104 (47 Tuc), allowed Ortolani et al. (1995) and Zoccali et al. (2003) to conclude that the bulge and the Halo are of the same age and that there are no traces in the bulge of an intermediate-age population. These authors recognise however that an age difference of $\sim$ 2-3 Gyr in either direction is possible.

The clusters of intermediate metallicity ($\sim~ -0.8$~dex) are very
important since they can be used to trace the
border between halo and disk. In this context, NGC 6366
(R.A.(2000)$=17^h27^m44^s$.3, DEC(2000)$=-05^{\circ} 04'36''$ ; l=$18.41^{\circ}$,
b=$+16.04^{\circ}$)
is an interesting globular cluster since, on one hand, it is metal
rich (with several [Fe/H] determinations that range between -0.65 and
-0.99 Pike 1976, Johnson et al.
1982, Zinn \& West 1984, DaCosta \& Seitzer 1989, Da Costa \& Armandroff
1995); and, on the other hand, it has a large heliocentric velocity of
$-$123.2 $\pm$ 1.0 km/s,
which associates the cluster with the slowly rotating halo cluster
system (Da Costa \& Seitzer 1989), and which would make NGC 6366 the
most metal-rich member of this cluster
population. Its Galactic position near to the Galactic disk contributes
to its high reddening, with estimates ranging between $E(B-V)=
0.65$ and $0.80$ (Harris 1976, Da Costa \& Seitzer 1989, Harris 1993).
Other Halo clusters with high metallicities  ([Fe/H] $\geq -1.0$) in the list of
Armandroff (1989) are NGC~6171, NGC~6569 and NGC~6712. An analysis of their 
kinematics led Cudworth (1988) and Da Costa \& Seitzer (1989) to conclude that NGC~6712 is a 
halo cluster, while the other two cannot be classified unambiguously. Thus, it is likely that NGC~6712 is the second most metal-rich ([Fe/H] = $-1.01$) cluster of the halo population.

Photometric studies of globular clusters are of special interest because
they can be used to generate a color magnitude diagram (CMD) that
provides insight into the cluster
age and evolutionary stage.
The CMD of NGC 6366 displays a clear turn off point, a well developed
red giant branch, and a
very incipient horizontal branch (HB) (Pike 1976; Harris 1993, Alonso
et al. 1997).
The appearance of the CMD has triggered an interest in the age
determination of the cluster and highlighted its importance in 
the study of galactic dynamics. 

A few estimates of the age of NGC~6366 can be found in the literature,
and their dispersion confirms the difficulty of dating globular
clusters. Alonso et al. (1997) found
an age of 18$^{+2}_{-3}$ Gyrs by comparing the CMD to the isochrones of
Straniero \& Chieffi (1991), and concluded that it is older by 4-6 Gyrs
than other metal rich
clusters.
However, Rosenberg et al. (1999), using a homogeneous data base of 34
globular clusters
 and a differential approach relative to a group of coeval clusters of
13.2 Gyrs of age (Carretta et al. 2000), estimated an age of about 11.0
Gyrs and hence concluding that the cluster
belongs to a group of clusters
younger than the coeval group. Salaris \& Weiss (2002), also applied a
differential approach relative to a group of clusters whose ages are
believed to be well determined
(M15, M3, NGC 6171 and 47 Tuc) and
 made an age estimate of 9.4-9.6 $\pm$ 1.4 Gyr for NGC~6366, depending
on the metallicity adopted, i.e. $\sim$ 2 Gyr younger than the reference clusters. This result further supports the idea that
the cluster is young and it
belongs to a group of young clusters linked to the galactic disk. 

The lack of a developed HB in the CMD of NGC~6366 is the cause of the
lack of RR Lyrae stars in the cluster, and the existence of a solitary
known RR Lyrae variable in the cluster is
therefore not
surprising. This variable star was detected and studied by Sawyer Hogg
(1973), and subsequently by Pike (1976) and Harris (1993). The latter
aimed his $BV$ photometry with CCD
observations to search for blue stragglers and variable stars. He
reported 27 blue straggler candidates but found no variable stars.
However, the photometry method used (PSF fitting
directly to the images) does not allow precise measurements in crowded
fields, limiting the possibility for detecting smaller amplitude
variables (Harris 1993). 

Difference image analysis (DIA) is a powerful technique allowing
accurate PSF photometry of CCD images, even in very crowded fields
(Alard \& Lupton 1998; Alard 2000; Bramich 2008).
In the present study, we apply a new algorithm for difference image
analysis (Bramich 2008) to a set of $V$ and $R$ images of NGC 6366 in order
to search for new variable stars down to $V=19.5$ mag. The paper is
organized as follows: in section 2 we summarise the observations
performed and describe the
data reduction methodology. In section 3 we discuss our approach to
searching for variable stars. In section 4 we report on the new
variables found and discuss their
nature. In section 5 we briefly discuss the Fourier decomposition of the
light curve of the RR Lyrae star, V1, and its implications for the
metallicity and distance of NGC 6366. In
section 6 we present our conclusions.

\section{Observations and reductions}
\label{sec:ObserRed}

The observations employed in the present work were performed, using the Johnson $V$ and $R$ filters, on May 5, 6 , 2006 and May 23, August 4, 5, September 4, 5, 2007.
We used the 2.0m telescope of the Indian Astronomical Observatory (IAO) at Hanle, India, located at 4500m above sea level. The estimated seeing was $\sim$1 arcsec.
The detector was a 
Thompson CCD of 2048 $\times$ 2048 pixels with a pixel
scale of 0.17 arcsec/pix and a field of view 
of approximately $11. \times 11.$ arcmin$^2$. Our data set consists of 61 
images in the V and 61 images in the R filters.

Image data were calibrated via standard overscan, bias and flat-field
correction procedures, and difference image analysis (DIA)
was performed on the images with the aim of extracting high precision
time-series photometry in the crowded field of NGC~6366.
We used a pre-release version of the {\tt DANDIA} software for the DIA
(Bramich, in preparation), which employs a new algorithm for
determining the convolution kernel matching a pair of images of the same
field (Bramich 2008).

\begin{figure}[!t]
\begin{center}
\includegraphics[width=8.5cm,height=8.5cm]{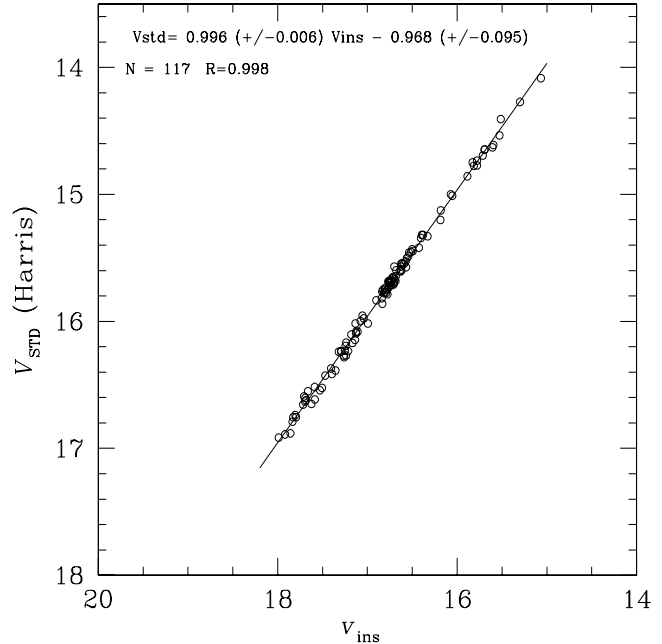}
\caption{The transformation relationship between the instrumental $v$ and the standard $V$ magnitudes.}
    \label{trans}
\end{center}
\end{figure}

Briefly summarising the DIA
procedure, we take the reference image for each filter as the
best-seeing image. We then measure fluxes (referred to as
reference fluxes) and positions for each PSF-like object (star) in the
reference image by extracting a spatially variable
(with polynomial degree 2) empirical PSF from the image and fitting this
PSF to each detected object. Deblending of
very close objects is attempted. Stars are matched between each image in
the sequence and the reference image, and
a linear transformation is derived which is used to register each image
with the reference image using cubic O-MOMS
resampling (Blu et al. 2001).

Each registered image is split into a 15 by 15 grid of subregions and a
set of kernels, modelled
as pixel arrays, are derived matching each image subregion to the
corresponding subregion in the reference image.
The kernel solution for each image pixel is determined by interpolating
the grid of kernel models using bilinear interpolation,
and the reference image, convolved with the appropriate kernel solution,
is subtracted from each registered image to
produce a sequence of difference images.

The differential fluxes for each star detected in the reference image
are measured
on each difference image as follows. The empirical PSF at the measured
position of the star on the reference image is determined
by shifting the empirical PSF model corresponding to the nearest pixel
by the appropriate sub-pixel shift using cubic O-MOMS resampling.
The empirical PSF model is then convolved with the kernel model
corresponding to the star position and current difference image.
Finally, it is optimally scaled to the difference image at the star
position using pixel variances $\sigma_{kij}^{2}$ for image $k$, pixel
column
$i$ and pixel row $j$, taken from the following standard CCD noise model:
\begin{equation}
\sigma_{kij}^{2} = \frac{\sigma_{0}^{2}}{F_{ij}^{2}} + \frac{M_{kij}}{G
F_{ij}}
\label{eqn:noise_model}
\end{equation}
where $\sigma_{0}^{2}$ is the CCD readout noise (ADU), $F_{ij}$ is the
master flat-field image, $G$ is the CCD gain (e$^{-}$/ADU) and
$M_{kij}$ is the image model (see Bramich 2008).

Lightcurves for each star are constructed by calculating the total flux
$f_{\mbox{\small tot}}(t)$ in ADU/s at each time $t$ from:
\begin{equation}
f_{\mbox{\small tot}}(t) = f_{\mbox{\small ref}} + \frac{f_{\mbox{\small
diff}}(t)}{p(t)},
\end{equation}
where $f_{\mbox{\small ref}}$ is the reference flux (ADU/s),
$f_{\mbox{\small diff}}(t)$ is the
differential flux (ADU/s) and $p(t)$ is the photometric scale factor (the
integral of the kernel solution).
Conversion to instrumental magnitudes is achieved using:
\begin{equation}
m(t) = 25.0 - 2.5 \log (f_{\mbox{\small tot}}(t)),
\end{equation}
where $m(t)$ is the magnitude of the star at time $t$. Uncertainties are
propagated in the correct analytical fashion.

\subsection{Transformation to the $V$ standard system}

The instrumental $v$ magnitudes were converted to the Johnson $V$
standard system by using the stars in the field of the cluster listed in
Table 2 of Harris (1993). Harris
reports $BV$ CCD magnitudes and colors tied to the photometric
system of Landolt (1973; 1983) for 315 stars. We have explored the instrumental
magnitudes of 160 of these stars contained in the field of our
collection of images and reject those stars with a root-mean-square
(RMS) scatter larger than 0.02 mag. In the end, we retained 117 stars
which we use as local standards.
Fig. \ref{trans} displays the relation between the instrumental and standard
magnitude systems which has the fitted form $V=(0.996\pm 0.006)~v~-~
(0.968 \pm 0.095)$. No significant color term was found.

The $r$ observations were retained in the instrumental system since no
standards with $R$ photometry in the field of the cluster were found in
the literature.

\begin{figure}[!t]
\begin{center}
\includegraphics[width=8.5cm,height=8.5cm]{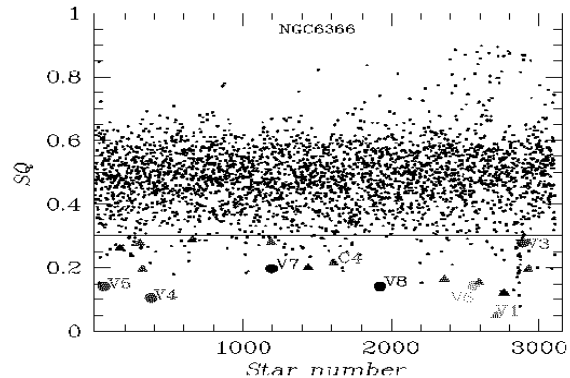}
\caption{$SQ$ parameter distribution of all stars measured in the field of NGC 6366. Stars below the $SQ = 0.3$ line are likely to be variables. The confirmed variables are labelled. V1 is the known RRab star. Red symbols are confirmed (dots) and suspected (triangles) long period variables. Likewise, blue symbols indicate confirmed and suspected eclipsing binaries. [$See ~the ~electronic 
~edition ~for ~a ~color ~version ~of ~this ~figure$]}
    \label{SQ}
\end{center}
\end{figure}

\section{New variables in NGC 6366}
\label{sec:newvar}

\subsection{Variable searching strategy}
\label{sec:strategy}

\begin{figure}[!t]
\begin{center}
\includegraphics[width=8.5cm,height=8.5cm]{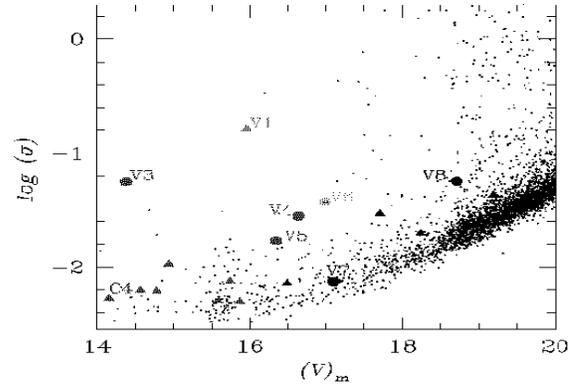}
\caption{Standard deviation as a function of the mean magnitude. Stars above the main
cluster of points are good candidates variables. The confirmed variables are labelled
as in Fig. \ref{SQ}. [$See ~the ~electronic 
~edition ~for ~a ~color ~version ~of ~this ~figure$]}
    \label{magerr}
\end{center}
\end{figure}

All the $V$ light curves of the 6172 stars measured in each of the 61 images available, were analyzed by the phase dispersion minimization approach (Burke et al. 1970; Dworetsky 1983). In this analysis the light curve is phased with the numerous test periods within a given range. For each period the dispersion parameter $SQ$ is calculated. When $SQ$ is at a minimum, the corresponding period is the best-fit period for that light curve. Bona fide variable stars should have a value of $SQ$ below a threshold. Similar analysis of clusters with numerous variables have shown that all periodic variables
are likely to have $SQ \leq 0.3$ (Arellano Ferro et al. 2004; 2006). Fig. \ref{SQ} shows the distribution of $SQ$ values for all stars measured in the images, with the threshold of $SQ=0.3$ indicated. Before proceeding 
 to explore the light curves of all stars with $SQ \leq 0.3$, the average magnitude and the standard deviation were calculated for each light curve. Fig. \ref{magerr} shows the dispersions (log $\sigma$) as a function of the magnitude-weighted mean $(V)_m$. Stars with a large dispersion for a given mean magnitude are, in principle, good variable candidates. However, it is possible that a light curve has a large $\sigma$ due to
occasional bad measurements of the corresponding star in some images, in
which case the variability is spurious.
We have used the $\sigma$ values to guide our search for variables in
the list of stars with $SQ \leq 0.3$, and, for promising candidate
variables, we checked the lightcurves by eye
for outliers, and we inspected the difference images to check for
possible contamination from nearby saturated stars and/or cosmetic
defects on the CCD. 

\begin{table}[hb]
\begin{center}
\caption{New variables in NGC 6366.}
   \label{newvar}
\begin{tabular}{ccccc}
\hline
\hline
Name &  $(V)_{m}$ & $(V-r)_{m}$ & P  & Type \\
&   &  & (days) & \\
\hline	
V3 &  14.29 & $-0.17$ & 3.72379 & AC? \\
\hline
V4 &   16.62 & +0.01 & --- & LPV \\
\hline
V5 &   16.31 & $-0.02$  & --- & LPV \\
\hline
V6 &   16.98 & $-0.30$ & 0.08018 & SX Phe$^1$\\
\hline
V7 &   17.09 & $-0.01$ & 0.475460 & LPV \\
\hline
V8 &   18.70 & +0.05 & 0.74291 &  E.B. \\
\hline
\hline
\end{tabular}
\end{center}
1. See Table \ref {SXFREQ}.
\end{table}

\begin{table}[hb!]

\begin{center}
\caption{Candidate variables in NGC 6366.}
   \label{possvarT}
\begin{tabular}{ccccc}
\hline
\hline
Name   & $(V)_{m}$ & $(V-r)_{m}$ & P & type  \\
&   &  & (days) & \\
\hline
C1  &  14.56 & +0.20 & --- & LPV \\
\hline
C2 &  14.78 & +0.10 & --- & LPV \\
\hline
C3 &   14.94 & +0.11 & --- & LPV \\
\hline
C4 &  14.15 & $-0.20$ & --- & AC? \\
\hline
C5 &   15.58 & +0.01 & --- & LPV \\
\hline
C6 &   15.86 & +0.29 & --- & LPV\\
\hline
C7 &   15.74 & +0.02 & --- & LPV\\
\hline
C8 &   16.48 & +0.14 & 1.14137 & E.B. ?\\
\hline
C9 &  17.69 & +0.03 & 0.86951 & E.B.  ? \\
\hline
C10 &  19.18 &$-0.24$ & 0.43171 & E.B.  ? \\
\hline
C11 &  18.23 &$-0.12$ & 0.57425 & E.B.  ? \\
\hline
\hline
\end{tabular}
\end{center}
\end{table}

\begin{figure}
\begin{center}
\includegraphics[width=8.5cm,height=8.0cm]{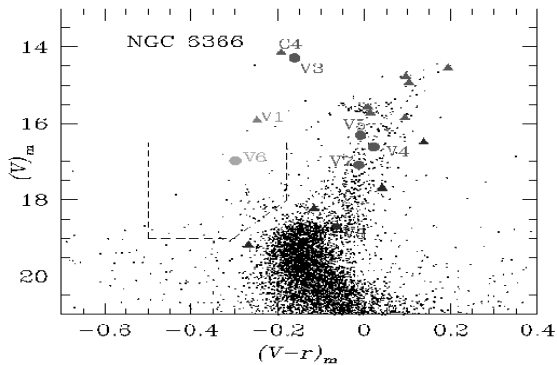}
\caption{CMD of NGC 6366 with the new confirmed and candidate variables indicated. The magnitudes and colours plotted are magnitude-weighted means over our entire collection of images. Symbols are as in Fig. \ref{SQ}. The $Blue~Stragglers$ region defined by Harris (1993) is enclosed by the dashed lines. [$See ~the ~electronic 
~edition ~for ~a ~color ~version ~of ~this ~figure$]}
   \label{CMD}
\end{center}
\end{figure}

The above procedure has allowed us to identify the group of variables
and candidate variables that are listed in Tables \ref{newvar} and
\ref{possvarT}, respectively. In these tables there are reported the 
magnitude-weighted mean magnitudes and colours $(V)_m$ and $(V-r)_m$,
which are also used to plot the stars in Figs. \ref{magerr} and \ref{CMD}.
We note that of the
previously known variables V1 and V2 in the cluster field, only V1 lies
within the field of view of our observations.

For variables with periods under 1 day, the period found by the phase
dispersion minimization method, produced a coherent light curve. In the
case of long term variables, the
sparse distribution of our observations (large gaps between observation
dates) does not allow a period to be estimated. However, it is clear
that these long term variables have
seasonal variations well above the photometric error in the lightcurve
measurements. 

The CMD of the cluster is shown in Fig. \ref{CMD} where the new confirmed and suspected variables are indicated. Individual stars are discussed in the following subsection. 

We recall at this point that the intensity-weighted mean  $<V>$ for variables reproduce better the magnitude for the equivalent static star than the magnitude-weighted mean $(V)_m$, that the difference 
 $(V)_m - <V>$ increases with the amplitude and that for symmetric light curves  $(V)_m = <V>$
(Bono et al. 1995). Of the variables discussed in this paper, V1 is the one with the largest amplitude 
(0.82 mag) and the most asymmetric light curve, and for this star $(V)_m - <V> \sim 0.1$. For the rest of the variables, the amplitudes are $\leq$ 0.1 mag and the variables in Table \ref{newvar} display symmetric light curves. Similar arguments hold for the colour $(V-r)_m$. Thus, we have chosen to plot $(V)_m$ and 
$(V-r)_m$ for all variables in Figs. \ref{magerr} and \ref{CMD}, trusting that the differences with $<V>$ and $<V-r>$ are negligible, particularly when identifying the types of the newly detected variables.

\subsection{New variables}
\label{sec:strategy}

\begin{figure}[!t]
\begin{center}
\includegraphics[width=7.5cm,height=9.5cm]{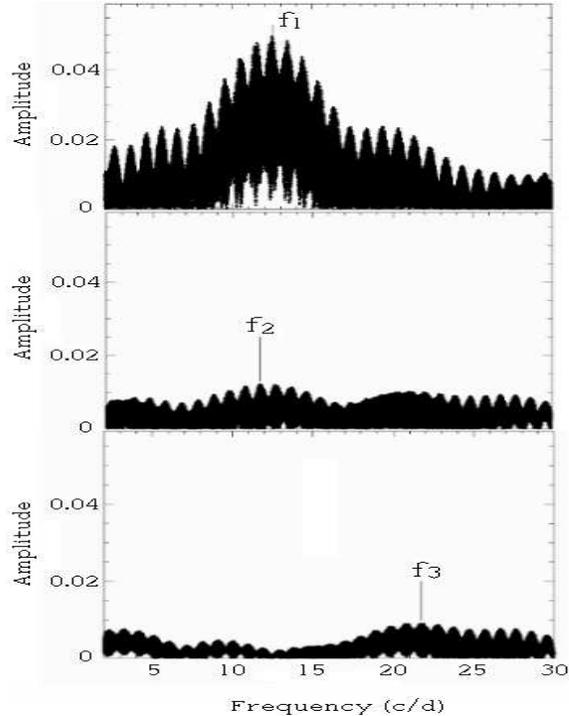}
\caption{Power spectrum of the SX Phe star, V6, showing the three active frequencies. The middle and bottom panels show the power spectra after the $f_1$ and $f_2$ frequencies have been respectively prewhitened.}
    \label{SXPHE}
\end{center}
\end{figure}

\begin{figure}[!t]
\begin{center}
\includegraphics[width=7.5cm,height=9.5cm]{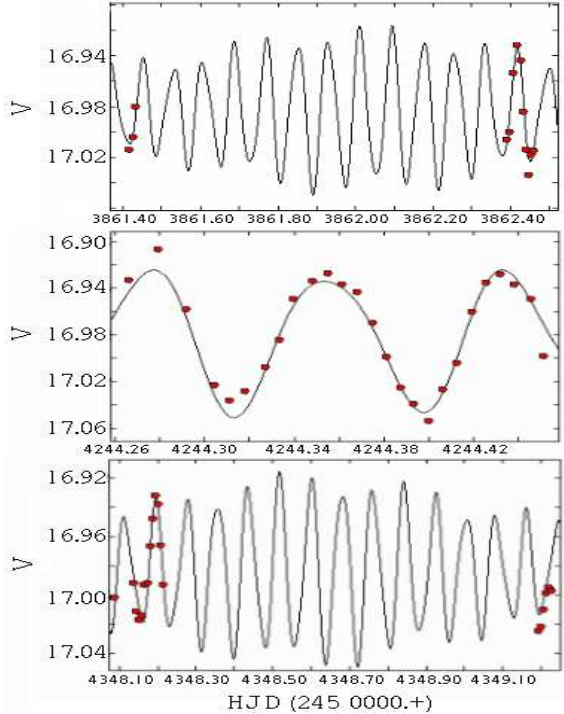}
\caption{Light curve of the SX Phe star V6, fitted by the superposition of the three active frequencies given in Table \ref{SXFREQ}.}
    \label{SXPHELC}
\end{center}
\end{figure}

{\bf V3.} This star shows a variation with an amplitude of $\sim$0.13 mag and a period of
3.72379 $\pm$ 0.00001 days. Its position on the CMD, $\sim$ 1.4 mag brighter than the RR Lyrae V1, suggest that this star might be an anomalous Cepheid (AC) or a Population II Cepheid (PC2). While ACs are common in dwarf spheroidal galaxies in the Local Group, they are rare in galactic globular clusters, only four are know;
V19 in NGC~5466 and three in $\omega~Cen$ (Clement et al. 2001). ACs are more massive than RR Lyrae stars, with masses  $\sim$ 1.8 to 4 $M_\odot$ (Caputo et al. 2004) and with periods between 0.5 and 3 days
(Wallerstein \& Cox 1984). These ranges of mass and period include the classical short-period Cepheids, which has led Dolphin et al. (2002) and Caputo et al. (2004) to suggest that rather than being a class of its own they are the natural extension to low metallicities of classical Cepheids. PC2s, on the other hand, are less massive than  RR Lyrae stars and have periods between 1 and 25 days, but they are not expected in 
globular clusters without blue HB (Smith \& Wehlau 1985). One exceptions seems to be the PC2 star in Palomar 3, a globular cluster with a red HB (Borissova, et al. 2000). This situation makes the PC2 classification also unlikely.
A distinct difference between ACs and PC2s is the shape of their light curves. The former have smaller amplitudes and more symetric light curves (Sandage \& Tammann 2006).
Although the period found for V3 in this work is marginally larger than in other ACs, its amplitude is smaller and light curve is more symmetrical than in PC2s. May the AC or PC2 classifications be unlikely, one cannot initially ruled them out. Our observations are, however,  too scanty to pin down the classification of this star and more observations are needed. It should be noted that, at this point, the possibility that V3 (and see below for C4) is not a member of NGC~6366
cannot be discarded. 

{\bf V4 and V5.} These stars display a long term variation in $V$ and $r$. They are long period variables sitting on the RGB.

{\bf V6.} This is a multiperiodic short period variable of the SX Phe type. Three frequencies were identified. These are listed in Table \ref{SXFREQ} along with their corresponding amplitudes and mode identification. Fig. \ref{SXPHE} shows the frequency spectra with the identification of the three modes. In Fig. \ref{SXPHELC}
the light curve of the SX Phe star is shown, fitted with the combination of the three identified modes.

The amplitudes of the modes are comparable to those identified in SX Phe stars in other globular clusters, for instance in NGC 5466 (Jeon et al. 2004). Similar to SX Phe stars in NGC~5466, V6 falls
on the $Blue~Stragglers$ region on the CMD defined for NGC~6366 by Harris (1993).

It is known that there is a Period-Luminosity ($P-L$) relationship for SX Phe stars
(McNamara 1995) which is difficult to determine due to the common mixture of modes
in these stars (McNamara 2001; Jeon et al. 2003). Linear $P-L$ relations independent of the metallicity can be found in the literature calculated in clusters of different metallicities, the slopes range between -3.25 and -1.62
(e.g. M53, -3.01, Jeon et al, 2003; NGC 5466, -3.25, Jeon et al, 2004; M55, -2.88 Pych et al. (2001); $\omega$ Cen, -1.62 McNamara 2000). 

The calibration of
Jeon et al. (2004) derived from the fundamental mode of seven SX Phe stars in NGC 5466 is of the form:

\begin{equation}
  \label{PL}
M_V = -3.25 (\pm 0.46) ~log P - 1.30 (\pm 0.06) ~~~~~~(\sigma=0.04).
\end{equation}

\noindent

These authors have discussed the value of the slope of eq. \ref {PL}
obtained from different clusters and have shown that the value $-3.25$
agrees within the uncertainties
with the above empirical determinations and with the theoretical predictions  ($-3.04$, Santolamazza et al. 2001; $-3.05$, Templeton et al. 2002).

Regarding the zero point of eq. \ref {PL}, Jeon et al. (2004) adopted  a true distance modulus of 16.0 for NGC~5466. It has been discussed by Arellano Ferro et al. (2008) that the true distance modulus of NGC~5466, in a scale where  the true distance modulus for the LMC is 18.5 $\pm$0.1 (Freddman et al. 2001; van den Marel et al. 2002; Clementini et al 2003), is  16.01 $\pm$ 0.09. Therefore we can say that eq. \ref {PL} produces distances consistent with the above mentioned true distance modulus of the LMC.
If we adopt
eq. \ref {PL} for the SX Phe in NGC~6366, and
$E(B-V) = 0.80$ (Harris 1993), we find a distance of $2.7 \pm 0.1$ kpc.

On the other hand, Nemec et al. (1994) proposed a $P-L$-[Fe/H] calibration for the fundamental mode of the form:

\begin{equation}
  \label{PLFE}
M_V = -2.56 (\pm 0.54) ~log P + 0.36 + 0.32{\rm [Fe/H]}.
\end{equation}

Which in turn, for an adopted value of [Fe/H] = -0.87, predicts a distance of $2.0 \pm 0.5$ kpc for the SX Phe in NGC~6366.
This estimation of the distance to NGC~6366 would be the shortest known in the literature. We shall further discuss the distance to the cluster in section 
4.2 in the light of the results derived for the RR Lyrae star V1.

\begin{figure*}[!t]
\begin{center}
\includegraphics[width=14.0cm,height=8.0cm]{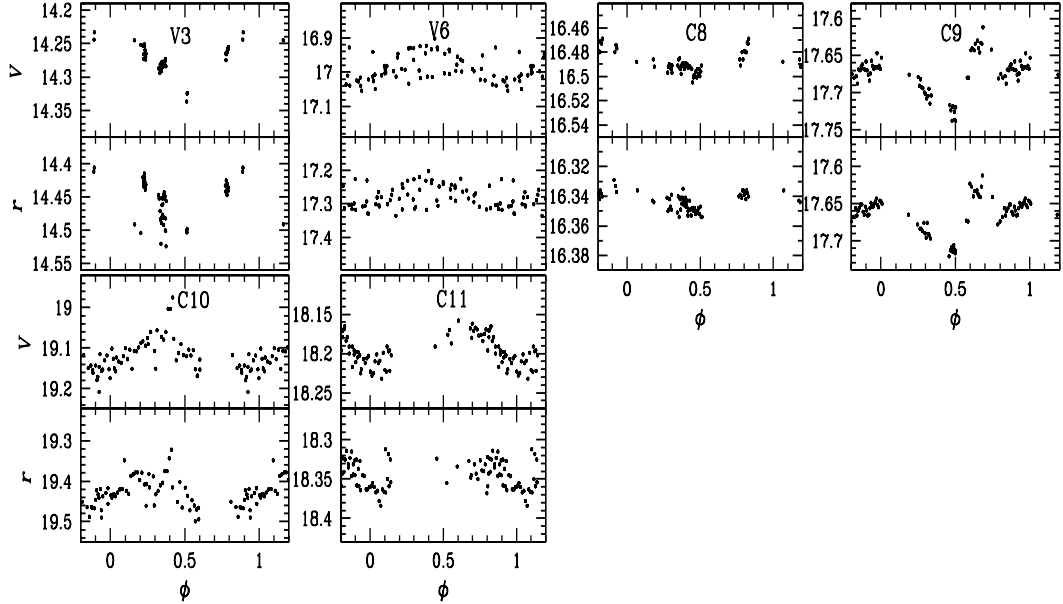}
\caption{Phased $V$ and $r$ light curves of short period confirmed and candidate variables.}
    \label{shortp}
\end{center}
\end{figure*}

\begin{table}[!b]
\begin{center}
\caption{Active modes in the SX Phe type star V6.}
   \label{SXFREQ}
\begin{tabular}{ccccc}
\hline
\hline
 &Frequency & Amplitude & mode & remark \\
& (c/d) & (mag) & &  \\
\hline	
$f_1$ &  12.4719 & 0.047 & $F$ &  \\
\hline
$f_2$ &  11.6469 & 0.015 & Nonradial &  \\
\hline
$f_3$ &  21.7118 & 0.009 & $2H?$ & $f_1/f_3=0.574$\\
\hline
\hline
\end{tabular}
\end{center}
\end{table}

{\bf V7.} This star was found about 0.02 mag dimmer on the first two nights than in the rest of the run. We consider it to be a long term variable whose periodicity cannot be estimated with the present data. There is the possibility that this variation corresponds to an eclipse, but more observations are required to confirm or refute this hypothesis.

{\bf V8.} This is an eclipsing binary with its eclipse clearly seen in the 
$V$ and $r$ light curves. The depth of the eclipse is about 0.15 mag. We have observed only a partial eclipse event and therefore the period cannot be accurately determined.

\subsection{Candidate variables}
\label{sec:possvarS}

{\bf C1, C2, C3, C4, C5, C6 and C7.} These are all possible long term variables
with seasonal variations larger than 0.02 mag in both $V$ and $r$ filters. The distribution of our observations, designed to detect short variations over a few hours,
does not facilitate further comment on their variable nature or an estimation of
their characteristic times and/or periods. 
More observations would be required to estimate a period and/or the
nature of variables.
C4, judged by its position on the CMD, like V3, could possibly be an AC or a PC2.
The limitations to accurately classify C4 at this point are like those discussed above for V3. Further observations are required to determine its amplitude, period, light curve shape, and membership in NGC~6366.

{\bf C8, C9, C10 and C11.}  On the other hand, these four stars display short period variations. Their periods, light curve shapes and 
position on the CMD suggest that these stars could be eclipsing binaries of the W UMa
type.

\begin{figure*}[!t]
\begin{center}
\includegraphics[width=17.0cm,height=16.0cm]{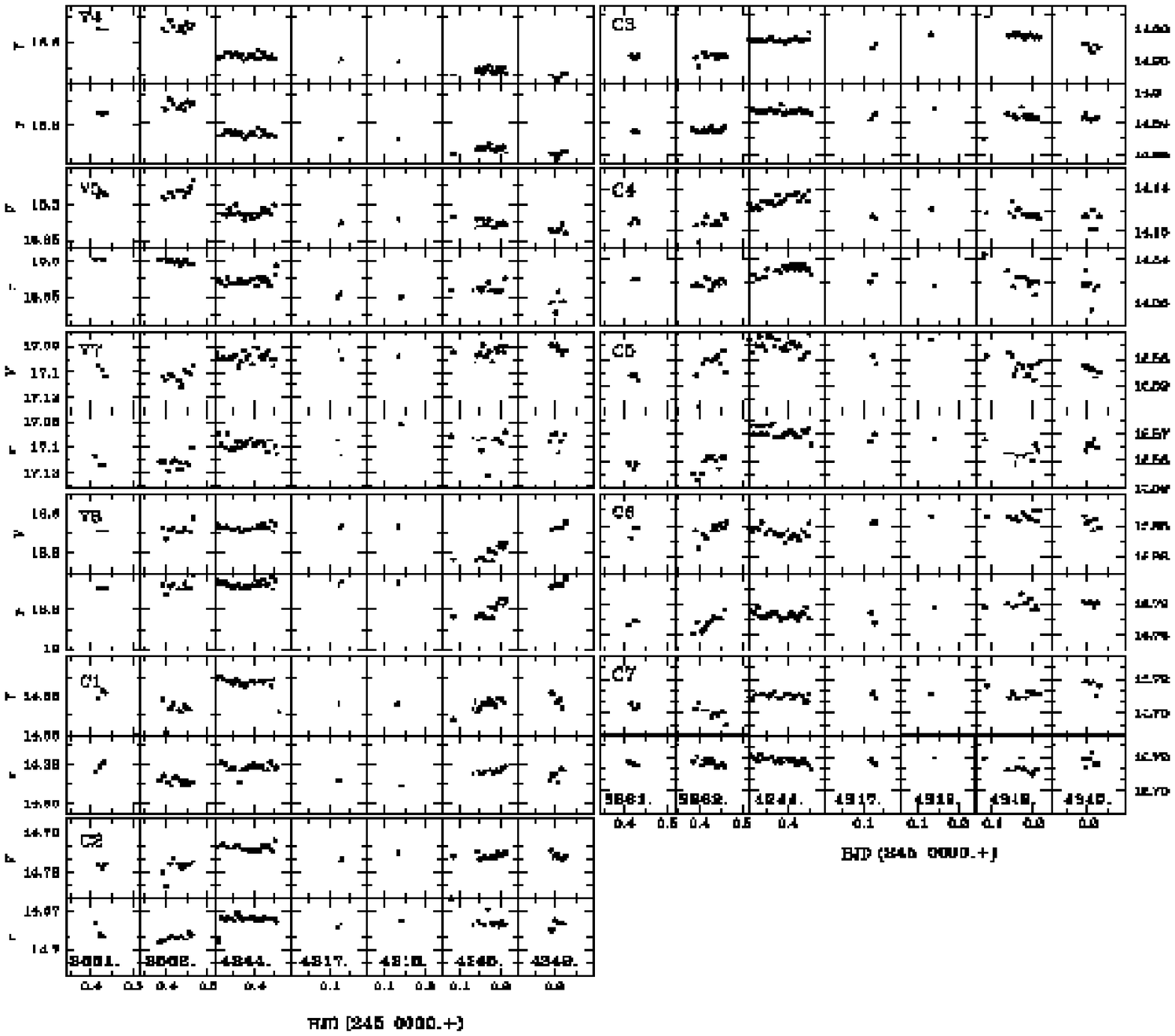}
\caption{$V$ and $r$ long term variations of confirmed and candidate variables. The horizontal axis is in julian day fraction. The integer number of each date is given inside the boxes at the bottom. The vertical scale is the same for the $V$ and $r$ light curves but it may be different from star to star.}
    \label{longp}
\end{center}
\end{figure*}

The light curves of both the short and the long period variables, in $V$ and $r$, are shown in Figs.
\ref{shortp} and \ref{longp}. The similarities of  the $V$ and $r$ light curve shapes  
 is yet another test of consistency and serves to confirm the variable nature of the star. All the variables in the field of the cluster are identified in Fig. \ref {elcumulo}

\begin{figure}[!b]
\begin{center}
\includegraphics[width=8.0cm,height=8.0cm]{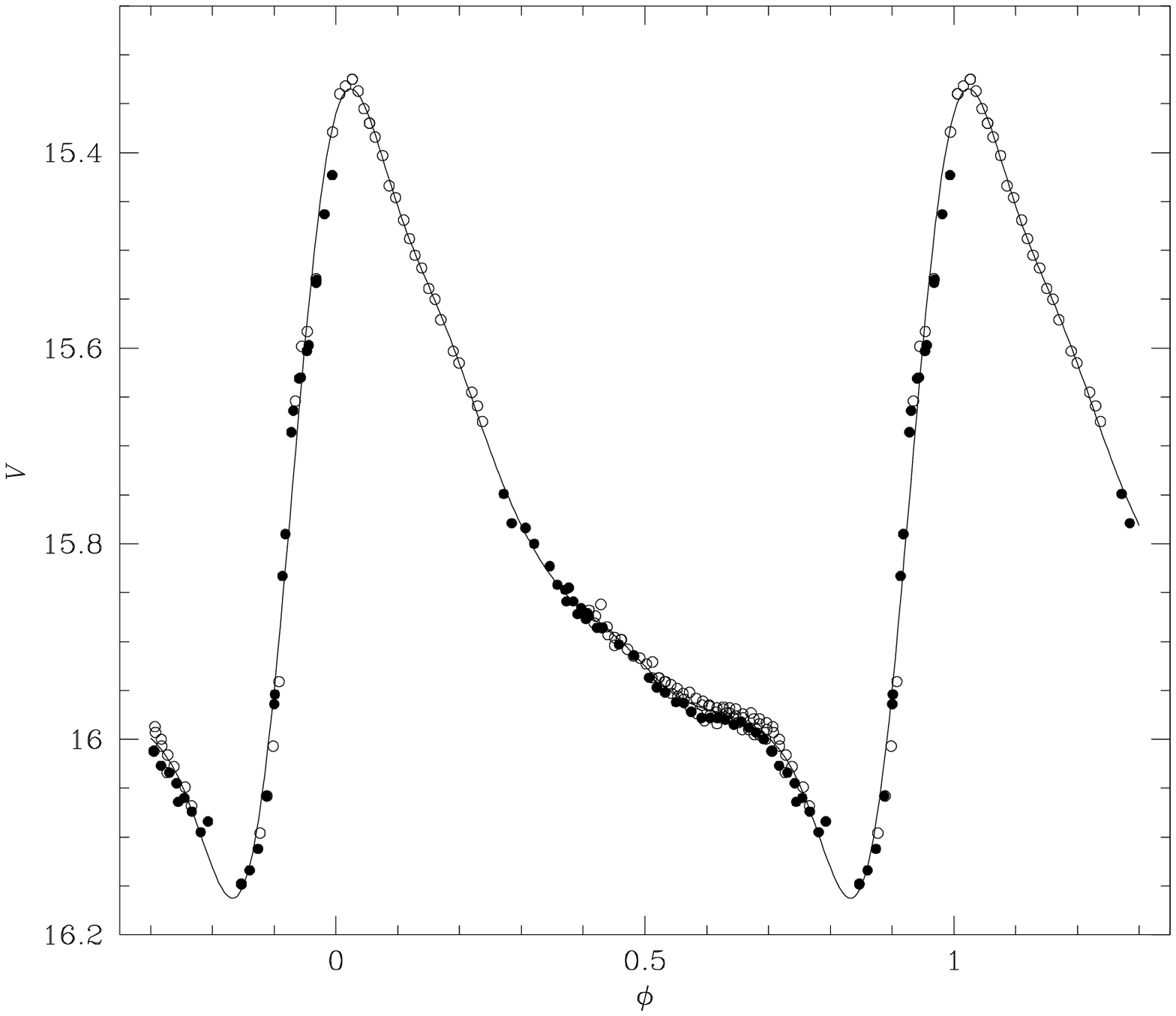}
\caption{Light curve of the RRab star V1. Dots are from the present work. Open circles are the observations of Harris (1993). The data are phased with a period of 0.5131635 days and an epoch of 244 7751.48. See text for discussion.}
    \label{RRLYR}
\end{center}
\end{figure}

\begin{table*}
\begin{center}
\caption{Fourier fit parameters and their uncertainties for the V light curve of the RRab star V1. \newline}
\label{foupar}
\begin{tabular}{cccccccc}
\hline
$  A_0$ & $A_1$ & $A_2$ & $A_3$ & $A_4$ & $A_5$ & $A_6$ &   \\
\hline
15.806 & 0.284 & 0.160 & 0.098 & 0.053 & 0.023 & 0.009   \\
0.005 & 0.002 & 0.003 & 0.005 & 0.005 & 0.002 & 0.003   \\
\hline
        & $\phi_1$ & $\phi_2$  & $\phi_3$ & $\phi_4$ &  $\phi_5$  & $\phi_6$  \\
\hline
      & 2.392 & 2.657 & 3.313 & 4.047 & 4.854 & 5.336  \\
      & 0.020 & 0.034 & 0.043 & 0.080 & 0.157 & 0.296  \\
\hline
 &  $\phi^{(c)}_{21}$ & $\phi^{(c)}_{31}$& $\phi^{(c)}_{41}$& &&&\\
\hline
&  4.156  & 2.419 & 0.761 & &&&\\
&  0.052  & 0.073 & 0.113 & &&&\\
\hline
\hline
\end{tabular}
\end{center}
\end{table*}

\section{The RR Lyrae star V1}
\label{sec:RRLV1}

The only RR Lyrae known in NGC 6366 is V1. Our V light curve was compared with that of Harris (1993) and found it to be  $\sim$0.02 mag fainter. Given the uncertainties in both photometries the agreement is reasonable.  We shifted our data to the magnitude level of Harris (1993) light curve. After this we used the two data sets to perform a period search and found a period of
0.5131635 $\pm$ 0.0000002 days, which is in excelent agreement to the final period used by Harris (1993) of 0.5131634 days to fit Pike's (1976) data taken in 1974 and his own from 1990. This confirms that the period has remained constant from 1974 to 2007. The two data sets phased with the period 0.5131635 days and an epoch of 244 7751.48 are shown in Fig.~\ref{RRLYR}. The full amplitude of the light curve is $A_V = 0.82$ mag. 

The position of V1 on the DCM is about 0.3 mag fainter than the more densely populated red edge of the HB, as it has also been noted by Harris (1993). According to Harris, the location of the star near the cluster center and its magnitude on the HB make its membership in NGC~6366 very likely. While evolved RR Lyrae stars are expected to be brighter than the red HB, ZAHB models of Brocato et al. (1999) do show that for a mixing length parameter $\alpha$ of 1.0, the distribution of stellar masses allows red HBs brighter than blue HB tails. 
However this low value of $\alpha$ would also produce a very blue red HB (e.g. Fig. 3 of Brocato et al. 1999), which is not observed in globular clusters. Ferraro et al. (2006) have shown that 
$\alpha$ is not significantly dependent on the metallicity and that a value of 2.17 is unique
for all globular clusters. This implies that RR Lyrae stars fainter than the red HB by as much as 0.3 mag cannot be produced by invoking a rather inefficient convection transport
in red giants, i.e. $\alpha \sim 1.0$.

Another RR Lyrae in a metal rich cluster is V9 in 47 Tuc ([Fe/H]=$-0.76$; Harris 1996). This star, in contrast with V1 in NGC~6366, is much hotter and it is brighter than the red HB and has been interpreted by Carney et al. (1993) as highly evolved. Harris (1993) has shown that the RGB and the red HB fiducial sequences in 47 Tuc and NGC~6366 match very well. Since $\alpha \sim 2.7$
for 47 Tuc (Ferraro et al. 2006), in NGC~6366 $\alpha$ must have a similar value, which is yet another argument against the low value of $\alpha$
in NGC~6366 and hence against the possibility of the formation of an underluminous RR Lyrae.
According to Carney et al. (1993) V9 has a much longer period for its blue amplitude $A_B$ and when plotted on the log~$P-A_B$ plane (Jones et al. 1992) there is indication that it is an evolved star. We do not have a $B$ light curve for V1 but, if the $V$ and $B$ amplitude ratio is similar to that in V9, for $A_V = 0.82$ one can forsee $A_B \sim 0.97$ for V1. Since log~$P = -0.2897$, it can be shown by plotting V1 in the log~$P-A_B$ plane, that it is consistent with metal rich ([Fe/H] $\geq \sim -0.6$) field 
RR Lyraes (see Fig. 12 of Jones et al. 1992). Therefore, the fact that V1 is $\sim 0.3$ mag fainter than the red HB and the above given arguments against a low value of the mixing length parameter $\alpha$ in NGC~6366, seem to lead to the conclusion that V1 does not reside in the cluster but rather beyond. In the following sections the metallicity and distance to V1 will be 
estimated and compared with the generally accepted values for NGC~6366. 

\begin{figure*}[!t]
\begin{center}
\includegraphics[width=14.cm,height=11.cm]{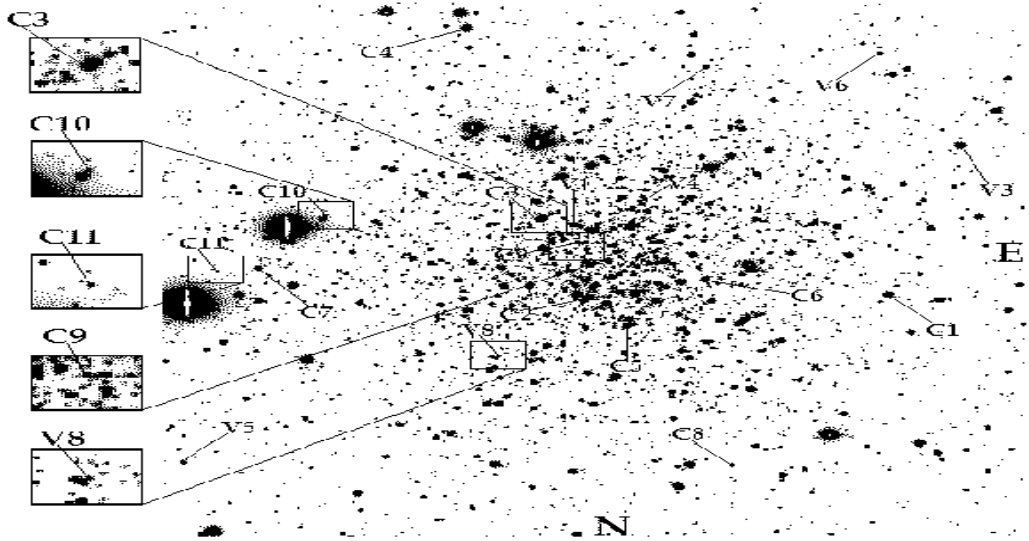}
\caption{Field of NGC 6366 with the new and candidate variables indicated. V1 is the known RRab star. Other names are as in Tables \ref{newvar} and 
\ref{possvarT}. A few stars with ambiguous identification are expanded in the left. The image is taken at the 2.0-m HCT of the IAO and it is approximately 10 $\times$ 10 arcmin$^2$.}
    \label{elcumulo}
\end{center}
\end{figure*}

\subsection{Fourier Decomposition of {\rm V1}}
\label{sec:Fourier}

 The mathematical representation of the light curve of V1 is of the form:

\begin{equation}
m(t) = A_o ~+~ \sum_{k=1}^{N}{A_k ~cos~( {2\pi \over P}~k~(t-E) ~+~ \phi_k ) },
\end{equation}

\noindent
where $m(t)$ are magnitudes at time $t$, $P$ the period and $E$ the epoch. A linear
minimization routine is used to fit the data with the Fourier series model, deriving
the best fit values of $E$ and of the amplitudes $A_k$ and phases $\phi_k$ of the sinusoidal components.

From the amplitudes and phases of the harmonics in eq. 4, the Fourier parameters, 
defined as $\phi_{ij} = j\phi_{i} - i\phi_{j}$, and $R_{ij} = A_{i}/A_{j}$, 
were calculated. The solid curve in Fig.~\ref{RRLYR} is the combination of 6 harmonics with amplitudes and phases $A_k$ and $\phi_k$ as listed in Table \ref {foupar}. The cosine  
$\phi^{(c)}_{21}$, $\phi^{(c)}_{31}$ and $\phi^{(c)}_{41}$ are also listed.

\subsection{On the metallicity and distance of V1 and NGC~6366}
\label{sec:PhysPar}

Calibrations of the iron abundance and absolute magnitude in terms of the Fourier parameters 
for RRab stars have been offered by Jurcsik \& Kov\'acs (1996) and
Kov\'acs \& Walker (2001) respectively. These calibrations are of the form;

\begin{equation}
\label{JK}
	{\rm [Fe/H]}_{\rm J} = -5.038 ~-~ 5.394~P ~+~ 1.345~\phi^{(s)}_{31},
\end{equation}

and

\begin{equation}
\label{KW}
M_V(K) = ~-1.876~log~P ~-1.158~A_1 ~+0.821~A_3 +K.
\end{equation}

\noindent
The standard deviations in the above equations are 0.14 dex and 0.04 mag respectively. In eq. \ref{JK}, the phase $\phi^{(s)}_{31}$ is calculated from a sine series. To convert the cosine series based $\phi^{(c)}_{jk}$ into the sine series $\phi^{(s)}_{jk}$,
one can use ~$\phi^{(s)}_{jk} = \phi^{(c)}_{jk} - (j - k) {\pi \over 2}$.

The metallicity [Fe/H]$_{\rm J}$ from eq. \ref{JK} can be converted to the 
metallicity scale of Zinn \& West (1984) (ZW) via [Fe/H]$_{\rm J}$ = 1.43 [Fe/H]$_{\rm ZW}$ + 0.88 (Jurcsik 1995). We used the period 0.5131635 days and the value of $\phi^{(c)}_{31}$ in Table \ref {foupar} duly transformed into $\phi^{(s)}_{31}$, to calculate 
[Fe/H]$_{\rm J} = -0.36$ which translates into [Fe/H]$_{\rm ZW} = -0.87$.

The zero point of eq.~\ref{KW}, K=0.43, has been calculated by Kinman (2002) using the prototype star RR Lyrae as calibrator, adopting for RR Lyrae the absolute magnitude $M_V= 0.61 \pm 0.10$ mag, as derived by Benedict et al.  (2002) using the star parallax measured by the HST. 
Kinman (2002) finds his result to be consistent with the coefficients of the $M_V$-[Fe/H] relationship given by Chaboyer (1999) and Cacciari (2003). All these results are consistent with the distance modulus of the LMC of $18.5 \pm 0.1$ (Freedman et al. 2001; van den Marel et al. 2002; Clementini et al. 2003).
The referee has led us to the recent paper by Catelan \& Cort\'es (2008) where these authors argue that the prototype RR Lyr has an overluminosity due to evolution of 0.064 $\pm$ 0.013 mag relative to HB RR Lyrae stars of similar metallicity. This would have to be taken into account if RR Lyr is used
as a calibrator of the constant $K$ in eq. \ref{JK}. While Catelan \& Cort\'es (2008) determined $M_V = 0.600 \pm 0.126$ for RR Lyr, i.e. very similar to the value quoted above from Benedict et al.  (2002), following Kinman's (2002) steps we find a new value of $K = 0.487$. Cort\'es \& Catelan (2008) have shown how the oveluminosity is also a function of the metallicity and have calibrated their equations in terms of the Str\"omgren color $c_{\rm o}$. Since we do not have $c_{\rm o}$ data for V1 in NGC~6366, the metallicity effect cannot be quantified but
we note that changing the value of $K$ between 0.43 and 0.487 produces a minor change in the derived distance to V1 from 3.24 to 3.16 kpc respectively. For the sake of homogeneity and better comparison with previous results (e.g. Arellano Ferro et al. 2008), in what follows we have adopted $K=0.43$.

Eqs.~\ref{JK} and \ref{KW} can be applied to light curves with a $compatibility ~  condition ~ parameter$ $D_m \leq$ 3. For the definition of $D_m$ see the works of Jurcsik \& Kov\'acs (1996) and  Kov\'acs \& Kanbur (1998). For the 6 harmonic light curve fit represented in Fig.~\ref{RRLYR} we find $D_m$=3.2, i.e. only marginally larger that the prescribed limit. If this criterion is applied to the 6 harmonic fit exclusively performed on Harris (1993) data, which has a better phase coverage, we find $D_m$=0.7. For consistency, the Fourier coefficients and physical parameters reported in Tables \ref{foupar} and \ref{fisicosAB} respectively, correspond to the calculated by fitting the data of Harris (1993) exclusively. It should be pointed however, that if the fit to both data sets had been used, the derived physical parameter would change well within
the quoted uncertainties.

\begin{table}
\begin{center}
\caption{\small Iron abundance, luminosity and distance estimates of the RRab star V1 from the Fourier light curve decomposition.}
\label{fisicosAB}
\hspace{0.01cm}
 \begin{tabular}{ll|ll}
\hline
[Fe/H]$_{\rm ZW}$ &  $-0.87 \pm 0.14$&  $\mu_o$& $12.55\pm 0.04$ \\
$M_V(K)$ &$0.68\pm 0.04$ &  $d(kpc)$& $3.2\pm 0.1$ \\
log$(L/L_{\odot})$&$1.623\pm 0.013$& $E(B-V)$ &$0.80^1$\\
\hline
\hline
\end{tabular}
\end{center}
1. Adopted from the value of Harris (1993) for NGC~6366.
\end{table}

The results for the iron abundance and luminosity for the RRab star, V1, are summarized in  Table \ref{fisicosAB} together with the distance $d$, and the true distance modulus $\mu_o$.

The value of [Fe/H]$_{\rm ZW}=-0.87 \pm 0.14$  found above for V1 can be compared  with previous estimations from different approaches for NGC~6366; Da Costa \& Seitzer (1989) found [Fe/H]$=-0.85~\pm~0.10$ using the strength of the  Ca II triplet at $\lambda\lambda~8498, 8542, 8662 \AA$ as a metallicity indicator in four giant stars in the cluster. Rutledge et al. (1997), also using the Ca II triplet, and transforming the metallicity to the scales of ZW and of Carretta \& Gratton (1997) (CG), found [Fe/H]$_{\rm ZW}=-0.58 \pm 0.14$ and [Fe/H]$_{\rm CG}=-0.73 \pm 0.05$, respectively. 

The value of $M_V(K) = 0.68 \pm 0.04$ for V1 and the adoption of $E(B-V) = 0.80$ and $R = 3.2$, lead to the distance $d({\rm V1}) = 3.2 \pm 0.1$ kpc. If, as discussed in section 4, V1 is beyond
NGC~6366 such that the star appears $\sim 0.3$ mag fainter than the cluster's red HB, we can still
estimate the distance to the cluster by shifting V1 0.3 mag to roughly the ZAHB, and then find 
$d({\rm NGC6366}) = 2.8 \pm 0.1$ kpc. Since V1 may also be evolved above the ZAHB, this distance estimate to NGC~6366 should be considered an upper limit.

On the other hand, referring to section 3.2, the distance $d({\rm V6}) = 2.7 \pm 0.1$ kpc was obtained
from the SX Phe
variable star V6 by considering the $P-L$ relation for SX Phe stars (Jeon et al. 2004). Alternativelly the $P-L$-[Fe/H] calibration of Nemec et al. (1994) for the fundamental mode in SX Phe stars produced a distance of $d({\rm V6}) = 2.0 \pm 0.5$ kpc.
Other determinations of the distance include 3.0 kpc with $E(B-V) = 0.80$ (Harris 1993) and 2.8 kpc with $E(B-V) =$ 0.70 $\pm$ 0.04 mag (Alonso et al. 1997). 
These results seem to disfavor the use of the $P-L$-[Fe/H] calibration proposed by Nemec et al. (1994) for a cluster as metal rich as NGC~6366.

\subsection{On the $M_V$- {\rm [Fe/H]} relationship for RR Lyrae stars}

The relation between $M_V$ and [Fe/H] for RR Lyrae stars has been traditionally represented in a linear fashion as $M_V$ =$\alpha$ [Fe/H] + $\beta$ and numerous calibrations by different techniques exist in the literature. Recent and very complete summaries on the calibration of this equation can be found in the works of Chaboyer (1999), Cacciari \& Clementini (2003) and Sandage \& Tammann (2006).
Most recent theoretical HB models do predict however a non linear relation between 
$M_V$ and [Fe/H], and again, thorough revisions of the results obtained since 1990 are given by Cacciari \& Clementini (2003) and Sandage \& Tammann (2006).

\begin{figure}[!t]
\begin{center}
\includegraphics[width=8.5cm,height=8.5cm]{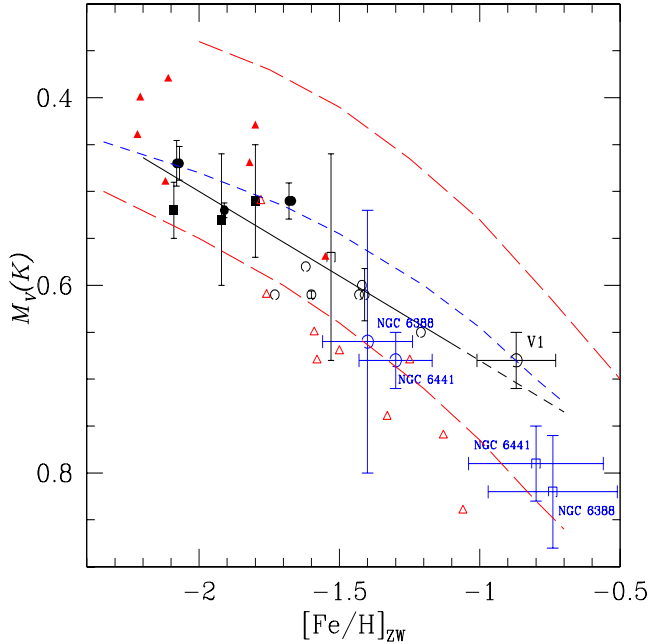}
\caption{The $M_V$-[Fe/H] relationship from the RR Lyrae Fourier light curve decomposition for a family of globular clusters taken from Arellano Ferro et al. (2008). Circles represent the results from the RRab stars while squares are from the RRc stars. NGC 6366 from the present work is labelled and its error bars correspond to 
dispersions of eqs. \ref{JK} and \ref{KW}. The best fit line has the form $M_V=+(0{\rm .}18\pm0{\rm .}03){\rm[Fe/H]}+(0{\rm .}85\pm0{\rm .}05)$. The cluster distribution of Caputo et al. (2000) are also included (triangles). The two metal-rich clusters NGC~6388 and NGC~6441 are included for completness in the high [Fe/H] domain.
Two predicted calibrations from the ZAHB models from Cassisi et al.
(1999) (top) and VandenBerg et al. (2000) (bottom) are shown as segmented lines.
The blue dashed line corresponds to the VandenBerg et al. (2000) ZAHB after the evolution
and its dependence on metallicity are considered.
Open symbols are used for OoI type clusters and filled symbols for OoII type clusters. See text in section 4.3 for detailed discussion. [$See ~the ~electronic 
~edition ~for ~a ~color ~version ~of ~this ~figure$]}
    \label{MVFEH}
\end{center}
\end{figure}

Recently Arellano Ferro et al. (2008) obtained a $M_V$-[Fe/H] relationship for clusters whose parameters have been estimated  by the Fourier decomposition of their RR Lyrae stars. These authors have converted the Fourier metallicities and absolute magnitudes to the ZW metallicity scale and to a distance scale where the LMC distance modulus is 18.5 $\pm$ 0.1 mag respectively.
The relationship found by these authors;
$M_V=+(0.18\pm0.03){\rm [Fe/H]}+(0.85\pm0.05)$, is reproduced in 
Fig.~\ref{MVFEH}, where the point corresponding to V1 has been added using the results in Table \ref{fisicosAB}. The error bars on V1 are from the dispersions of eqs. \ref{JK} and \ref{KW}. The position of V1 corresponds well with a linear extrapolation to the metal rich domain of the above linear $M_V$-[Fe/H] relationship. The solid line in  Fig.~\ref{MVFEH} is from Arellano Ferro et al. (2008)
(V1 not included in the fit) and it implies  $M_V(RR)$=0.58$\pm$0.05 mag for [Fe/H]=$-1.5$, in excellent agreement with independent calibrations
of Chaboyer (1999) (0.58$\pm$0.12) and Cacciari \& Clementini (2003) (0.59$\pm$0.03).

We have also included in Fig.~\ref{MVFEH} the cluster distribution calculated by Caputo et al. (2000) for [$\alpha$/Fe]~=~0.3 and which has often been used as an 
empirical evidence of the non-linearity of the $M_V$-[Fe/H] relationship (triangles). In fact Caputo and co-workers have suggested two linear fits for the metallicity domains separated at [Fe/H]=$-1.5$. Since Caputo et al. metallicities are given in the Carretta \& Gratton (1997) metallicity scale, to fairly compare with the Fourier
results we have converted their metallicities into the ZW metallicity scale (see Carretta \& Gratton 1997). There are also included two extreme and rather emblematic predicted calibrations for the ZAHB; by Cassisi et al. (1999) (upper segmented line) and by VandenBerg et al. (2000) for [$\alpha$/Fe]~=~0.3 (lower segmented line). Since during their HB evolution the stars spend most of their time at 0.1-0.2 mag brighter than the ZAHB (Cacciari \& Clementini 2003), to take into account evolutionary effects and to better compare with the empirical results, 
the $M_V(ZAHB)$-[Fe/H] theoretical relations must be shifted  to convert them  into $M_V(RR)$-[Fe/H]. Such a shift is a function of [Fe/H] and the difficulties in quantifying it have been amply discussed in the review by Gallart et al. (2005) (see their Fig. 9). We have adopted the Cassisi \& Salaris (1997) relation between
$M_V(ZAHB)$-$M_V(RR)$ and [Fe/H] to transform the predicted $M_V(ZAHB)$-[Fe/H] relationship from the
ZAHB models of VandenBerg et al. (2000) into its corresponding evolved $M_V(RR)$-[Fe/H] relationship (dashed blue line in 
Fig.~\ref{MVFEH}). This evolved $M_V(RR)$-[Fe/H] relationship is the one that should be compared with 
the empirical results obtained from RR Lyrae stars through the Fourier approach.

Given their internal accuracies and dispersions, the results from the Fourier solutions coincide with the independent determinations from Caputo et al. (2000), rather satisfactorily, particularly for the low metallicity domain. Also, and very significantly, the Fourier results agree, within the uncertainties, 
with the theoretical prediction 
from VandenBerg et al.'s (2000) ZAHB models and the evolutive effects and their dependence on the metallicity calculated by
Cassisi \& Salaris (1997). While it is true that the Fourier results seem to suggest a linear distribution, enhanced by the appearance 
of V1 after the results in the present paper, it would be very interesting to incorporate RR Lyrae stars in metal-rich clusters, such as NGC~6388 ([Fe/H]=$-0.60$) and NGC~ 6441 ([Fe/H]=$-0.53$), 
to the sample of Fourier analyzed systems to study the behaviour of the  $M_V(RR)$-[Fe/H] relationship
in the high metallicity range. The light curves of some RRab and RRc stars in NGC~6388 and NGC~6441 have been 
Fourier-decomposed by Pritzl et al. (2002, 2001) respectively. The metallicity values found by Pritzl and collaborators for the RRab stars in these clusters, already transformed to the ZW metallicity scale are: [Fe/H]$_{\rm ZW}^{6388}=-1.4 \pm 0.16$, [Fe/H]$_{\rm ZW}^{6441}=-1.3\pm 0.13$, $M_V^{6388}=0.66\pm 0.14$ and $M_V^{6441}=0.68\pm 0.03$. For the RRc stars they found $M_V^{6388}=0.82\pm 0.06$ and $M_V^{6441}=0.79\pm 0.03$. We have complemented the RRc star by estimating [Fe/H] by means of the calibration of Morgan et al. (2007) and adopting the Fourier parameters published by Pritzl and collaborators, the results are [Fe/H]$_{\rm ZW}^{6388}=-0.74 \pm -0.23$
and [Fe/H]$_{\rm ZW}^{6441}=-0.80 \pm 0.24$. The uncertainties are the standard deviations of the mean.
These results are plotted and labeled following the symbol conventions on Fig. \ref{MVFEH}. The points corresponding to NGC~6388 and NGC~6441 might lean toward favouring a non-linear $M_V(RR)$-[Fe/H] relationship, leaving V1 at an 
odd position. However a few words of caution are necessary. Pritzl et al. (2001, 2002)
noted already the low values of [Fe/H] for the RRab stars estimated from the Fourier approach when compared with the values by Armandroff \& Zinn (1988) from the Ca II IR triplet. Pritzl et al. have discussed the possibility of metallicity spread in NGC~6388. It is known that RR Lyraes in these two clusters have unusually long periods, which has originated the suggestion that they are of the OoII type, and it is uncertain whether the Jurcsik-Kov\'acs calibrations
are valid  in such case. For the  $M_V$ values of RRc stars, Pritzl et al. (2001, 2002) have used the Jurcsik (1998) calibration $M_V$(P,$A_1,\phi_{31}$) but one can use alternativelly 
the Kov\'acs (1998) calibration $M_V$(P,$A_4,\phi_{21}$). The zero point of the later calibration has been disputed by Cacciari et al. (2005) who have suggested that, for M3, the zero point should be decreased by $0.2 \pm 0.02$. This exercise produced distances for metal poor clusters (e.g. M3 and NGC~5466) which are consistent with the luminosities of RR Lyrae in the LMC and a distance modulus of LMC of $18.5 \pm 0.1$
(e.g. Cacciari et al 2005; Arellano Ferro et al. 2008).
The zero point of these calibrations might be metallicity dependent and it is not clear what the offset should be for metal richer clusters like NGC~6388 and NGC~6441. Given these considerations, the RRc points for NGC~6388 and NGC~6441 on Fig. \ref{MVFEH} might need to be shifted to brighter magnitudes by as much as 0.2 mag. Thus, the Fourier results on these two metal rich clusters cannot be given too much weight
in determining the shape of the $M_V(RR)$-[Fe/H] relationship in the high metallicity range.

The amplitude and period of V1, $A_V=0.82$ and log~$P = -0.29$, place the star on the $A_V-log P$ plane among the field metal-rich ([Fe/H] $\geq -0.8$) RR Lyraes, and not among the RR Lyraes of the metal rich cluster NGC~6388 ([Fe/H]=-0.6; Armandroff \& Zinn 1988) which have unusually large periods for a given amplitude (see for instance Fig. 9 of Pritzl et al. 2002). On the plane P$_{\rm ab}$-[Fe/H] the young and the old galactic cluster populations show the Oosterhoff dichotomy (see Fig. 5 of Catelan 2005) and likewise, the V1 is positioned
as an extension of the OoI group. If V1 was a member of NGC~6366, on the basis of the above comparisons, it would seem reasonable to consider NGC~6366 of the OoI type. However, since
V1 is likely a non-member of the cluster, no Oo type can formally be assigned to NGC~6366.

\section{Conclusions}
\label{sec:Concl}

The use of difference imaging has enabled us to perform precision
photometry in the globular cluster NGC~6366 and thereby lead to the detection of confirmed and possible new variables. The difference imaging technique used employs a new algorithm
for determining the convolution kernel as a pixel grid. Among the new
variables that we have found, we
have identified one SX Phe star with at least three active modes, two
possible AC's (or P2C's), one eclipsing binary and three long period red
variables. We have also detected
possible variations in a group of long term variables and short period
eclipsing binaries likely to be of the W~UMa type.

Despite the position of V1 very near the center of NGC~6366, the membership of V1 in the cluster 
is doubted mainly because the star is $\sim 0.3$ mag
fainter than the red HB in the CMD, and because an inneficient convection transport in the 
red giants in NGC~6366 cannot be invoked as a possible cause for a real underluminosity of V1.
Therefore, the metallicity and distance estimated for V1 from the Fourier technique, cannot be
considered as representative of the cluster. We note however that the metallicity found for  V1,
[Fe/H]$_{\rm ZW}=-0.87 \pm 0.14$, 
is very similar to values ascribed to NGC~6366 by independent spectroscopic estimates for giant stars in the cluster, (e.g [Fe/H]$_{\rm ZW}=-0.85~\pm~0.10$, Da Costa \& Seitzer 1989; [Fe/H]$_{\rm ZW}=-0.58 \pm 0.14$, Rutledge et al. 1997). 

 The distance to V1 was estimated as $d({\rm V1}) = 3.2 \pm 0.1$ kpc. If V1 is shifted 0.3 mag to the ZAHB, and considering that V1 might be evolved above the ZAHB, an upper limit for the distance to the cluster of $d({\rm NGC~6366}) = 2.8 \pm 0.1$ kpc can also be estimated. An independent
determination of the distance to NGC~6366 from the $P-L$ relationship for SX Phe stars and the pulsation modes identified in the SX Phe star V6 found in the cluster, gives the distance $d({\rm V6}) = 2.7 \pm 0.1$ kpc, which is consistent with the upper limit determined from V1.

The position of V1 on the $M_V$-[Fe/H] plane suggests a linear extrapolation to the metal-rich domain.
Inclusion of RR Lyrae stars in the metal rich clusters NGC~6388 and NGC~6441 seem to suppot the non-linear
behavior of the $M_V$-[Fe/H] relationship, however it must be stressed that those RR Lyraes have unusual long periods for their amplitudes and then, the Fourier decomposition calibrations to determine their [Fe/H] and $M_V$ values
may not be applicable.

The $M_V(RR)$-[Fe/H] relationship derived from the Fourier results compares well, within the uncertainties, with the clusters distribution from the analysis of Caputo et al. (2000) and with the
$M_V(ZAHB)$-[Fe/H] relationship prediction from ZAHB models of VandenBerg et al. (2000) once evolution from the ZAHB is considered. The excellent agreement of the position of V1 with this theoretical prediction
further supports the idea that V1 is an evolved object from the ZAHB and that its apparent underluminosity 
in the CMD is due to its non-membership in NGC~6366. This interpretation has to compete with the otherwise
straight one that, based on the similarity of the metallicities of the V1 and of the cluster and sitting
V1 so close to the center of the cluster, V1 is very likely a member of the cluster. In
this later case the observed underluminosity of V1 relative to the red HB is yet to be understood.

\bigskip

\noindent
We are grateful to the support astronomers of IAO, at Hanle and CREST (Hosakote), for their very efficient help while acquiring the data. AAF and VRL acknowledge support from DGAPA-UNAM grant through project IN108106. RF and PR are grateful to the CDCHT-ULA for finantial support through project C-1544-08-05-F. 
Very useful corrections, comments and suggestions made by an anonymous and very constructive
referee are deeply appreciated, particularly those
on the non-membership of V1 in NGC~6366. This work has made a large use of the SIMBAD and 
ADS services, for which we are thankful.


\begin{thebibliography}

\bibitem{} Alard, C., 2000, A\&AS, 144, 363.
\bibitem{} Alard, C., Lupton, R.H., 1998, ApJ, 503, 325.
\bibitem{} Alonso, A., Salaris, M., Mart\'inez-Roger, C., Straniero, O., Arribas, S., 1997, A\&A, 323, 374.
\bibitem{} Arellano Ferro A., Ar\'evalo M. J., L\'azaro C., Rey M., Bramich D. M., Giridhar S., 2004, RevMexAA, 40, 209
\bibitem{} Arellano Ferro, A., Garc\'ia Lugo, G., Rosenzweig, P., 2006, RevMexAA, 42, 75.
\bibitem{} Arellano Ferro, A., Rojas, L\'opez, V, Giridhar, S., Bramich, D.M., 2008, MNRAS, 384, 1444.
\bibitem{} Armandroff, T.E., 1989, AJ, 97, 375.
\bibitem{} Armandroff, T.E., Zinn, R, 1988, AJ, 96, 92.
\bibitem{} Benedict, G. F., McArthur, B. E., Fredrick, L. W., et al. 2002, ApJ, 123, 473. 
\bibitem{} Blu, T. The\'venaz, P., Unser, M., 2001, IEEE Trans. Image
Process., 10, 1069
\bibitem{} Bond, I. A., Abe F., Dodd R. J., 2001, MNRAS, 327, 868.
\bibitem{} Bono, G., Caputo, F., Stellingwerf, R. F., 1995, ApJS, 99, 263.
\bibitem{} Borissova, J., Ivanov, V.D., Catelan, M., 2000, IBVS, No. 4919.
\bibitem{} Bramich D. M., 2008, MNRAS, 386, L77.
\bibitem{} Bramich D. M., Horne K., Bond, I. A., Street, R. A., Cameron, A. C., Hood, B., Cooke, J., James, D., Lister, T. A., Mitchell, D., Pearson, K., Penny, A., Quirrenbach, A., Safizadeh, N., Tsapras, Y., 2005, MNRAS, 359, 1096.
\bibitem{} Brocato, E., Castellani, V., Raimondo, G., Walker, A.R., 1999, ApJ, 527, 230.
\bibitem{} Burke, E.W., Rolland, W.W., Boy, W.R., 1970, JRASC, 64, 353.
\bibitem{} Cacciari, C., 2003, in ASP Conf. Series 296, New Horizons in Globular
Clusters Astronomy, eds G. Piotto \& G. Meylan, (San Francisco; ASP), p. 329.
\bibitem{} Cacciari, C., Clementini, G. 2003, in Lectures Notes in Physics Vol. 635, Stellar Candles for Extragalactic Distance Scale. eds. Alloin, D., Gieren, W. Kluwer, Dordrecht, p. 105.
\bibitem{} Cacciari, C., Corwin, T.M., Carney, B.W., 2005, ApJ, 129, 267.
\bibitem{} Caputo, F., Castellani, V., Marconi, M., Ripepi, V., 2000, MNRAS, 316, 819.
\bibitem{} Caputo, F., Castellani, V., Degl'Innocenti, S., Fiorentino, G., Marconi, M., 2004, A\&A, 424, 927.
\bibitem{} Carney, B.W., Storm, J., Williams, C., 1993, PASP, 105, 294.
\bibitem{} Cassisi, S., Castellani, V., Degl'Innocenti, S., Salaris, M., Weiss, A., 1999, A\&AS, 134, 103.
\bibitem{} Cassisi, S., Salaris, M., 1997, MNRAS, 285, 593.
\bibitem{} Carretta E., Gratton R. G., 1997, A\&AS, 121, 95.
\bibitem{} Carretta E., Gratton R. G., Clementini G., Fuci Pecci F., 2000, ApJ, 533, 215.
\bibitem{} Catelan, M., 2005 in Resolved Stellar Populations, eds. Valls-Gabaud, D. \& Ch\'avez, M., ASP Conf Ser. in press (astro-ph/0507464).
\bibitem{} Catelan, M., Cort\'es, C., 2008, ApJ, 676, L135.
\bibitem{} Clement, C.M., Muzzin, A., Dufton, Q., Ponnampalam, T., Wang, J., Burford, J., Richardson, A., Rosebery, T., Rowe, J., Sawyer-Hogg, H., 2001, AJ, 122, 2587.
\bibitem{} Clementini, G.,  Gratton R. G., Bragaglia, A., et al. 2003, AJ, 125, 1309
\bibitem{} Chaboyer, B., 1999, in Post-Hipparcos Cosmic Candles, eds. A. Heck \& F. Caputo (Dordrech: Kluwer), p. 111
\bibitem{} Cort\'es, C., Catelan, M., 2008, ApJS, 177, 362.
\bibitem{} Cudworth, K.M., 1988, AJ, 96, 105.
\bibitem{} Da Costa, G.S., Seitzer, P., 1989, AJ, 97, 405.
\bibitem{} Da Costa, G.S., Armandroff, T.E., 1995, AJ, 109, 2533.
\bibitem{} Dolphin, A.E., Saha, A., Claver, J., Skillman, E.D., Cole, A. A., Gallagher, J.S., Tolstoy, E., Dohm-Palmer, R.C., Mateo, M., 2002, AJ, 123, 3154.
\bibitem{} Dworetsky, M.M., 1983, MNRAS, 203, 917.
\bibitem{} Ferraro, F. R., Messineo, M., Fusi Pecci, F., de Palo, M. A., Straniero, O., Chieffi, A., Limongi, M., 1999, 118, 1738.
\bibitem{} Ferraro, F. R., Valente, E., Srtaniero, O., Origlia, L., 2006, ApJ, 642, 225.
\bibitem{} Freedman, W.L., Madore, B.F., Gibson, B.K. et al. 2001, ApJ, 553, 47.
\bibitem{} Gallart, C., Zoccali,, M., Aparicio, A., 2005, Ann.Rev.A\&A., 43, 387.
\bibitem{} Harris, H.C., 1976, AJ, 81, 1095.
\bibitem{} Harris, H.C., 1993, AJ, 106, 604.
\bibitem{} Harris, W.E., 1996, AJ, 112, 1487.
\bibitem{} Jeon, Y.-B., Lee M.G., Kim S.-L, Lee H., 2003, AJ, 125, 3165.
\bibitem{} Jeon, Y.-B., Lee M.G., Kim S.-L, Lee H., 2004, AJ, 128, 287.
\bibitem{} Johnson, H.R., Mould, J., Bernardt, A.P., 1982, ApJ, 258, 261.
\bibitem{} Jones, R.V., Carney, B.W., Storm, J., Latham, D.W., 1992, ApJ, 386, 646.
\bibitem{} Jurcsik, J., Acta Astron., 1995, 45, 6653.
\bibitem{} Jurcsik, J., 1998, A\&A, 333, 571.
\bibitem{} Jurcsik, J., Kov\'acs G., 1996, A\&A, 312, 111.
\bibitem{} Kinman, T.D., 2002, Inf. Bull. Var. Stars, No. 535
\bibitem{} Kov\'acs, G., 1998, Mem. Soc. Astron. Ital., 69, 49.
\bibitem{} Kov\'acs, G., 2002, in ASP Conf. Ser. 265, $\omega$ Centauri: A Unique Window into Astrophysics. Eds. van Leeuwen, F., Hughes, J., Pioto, G., (San Francisco; ASP), p. 163
\bibitem{} Kov\'acs, G., Kanbur, S.M., 1998, MNRAS, 295, 834. 
\bibitem{} Kov\'acs, G., Walker, A.R., 2001, A\&A, 371, 579
\bibitem{} Landolt, A.U., 1973, AJ, 78, 959.
\bibitem{} Landolt, A.U., 1983, AJ, 88, 439.
\bibitem{} McNamara, D.H., 1995, AJ, 109, 1751.
\bibitem{} McNamara, D.H., 2000, PASP, 112, 1096.
\bibitem{} McNamara, D.H., 2001, PASP, 113, 335.
\bibitem{} Minniti, D., 1995, AJ, 109, 1663.
\bibitem{} Morgan, S.M., Wahl, J.N., Wieckhorst, R.M., 2007, MNRAS, 374, 1421. 
\bibitem{} Nemec, J.M., Linnell Nemec, A.F., Lutz, T.E., 1994, AJ, 108, 222.
\bibitem{} Ortolani, S., Renzini, A., Gilmozzi, R., Marconi, G., Barbuy, B., Bica, E., Rich, R. M., 1995, Nature, 377, 701.  
\bibitem{} Pike, C.D., 1976, MNRAS, 177, 257.
\bibitem{} Pritzl, B.J., Smith, H.A., Catelan, M., Sweigart, A.V., 2001, AJ, 122, 2600. 
\bibitem{} Pritzl, B.J., Smith, H.A., Catelan, M., Sweigart, A.V., 2002, AJ, 124, 949.
\bibitem{} Pych, W., Kaluzny, J., Krzeminski, W., Schwarzenberg-Czerny, A., Thompson, I. B., 2001, A\&A, 367, 148
\bibitem{} Rosenberg, A., Saviane, I., Piotto, G., Aparicio, A., 1999, AJ, 118, 2306.
\bibitem{} Rudledge, G.A., Hesser, J.E., Stetson, P.B., 1997, PASP, 109, 907.
\bibitem{} Salaris M., Weiss A., 2002, A\&A, 388, 492.
\bibitem{} Sandage, A., Tammann, G.A., Ann Rev. A\&A, 44, 93, 2006.
\bibitem{} Santolamazza, P., Marconi, M., Bono, G., Caputo, F., Cassisi, S., GIlliland, R.L., 2001, ApJ, 554, 1124.
\bibitem{} Sarajedini, A., 1997, AJ, 113, 682.
\bibitem{} Stetson, P.B., Bolte, M., Harris, W.E., Hesser, J.E., van den Bergh, S., VandenBerg, D.A., Bell, R.A., Johnson, J.A., Bond, H.E., Fullton, L.K., Fahlman, G.G., Richer, H.B., 1999, AJ, 117, 247 
\bibitem{} Sawyer Hogg, H.G., 1973, Pubs. David Dunlap Obs., 3, No. 6.
\bibitem{} Smith, H.A., Wehlau, A., 1985, ApJ, 298, 572
\bibitem{} Straniero, O., Chieffi, A., 1991, ApJS, 76, 525.
\bibitem{} Stetson, P.B., 1987, PASP, 99, 191
\bibitem{} Templeton, M., Basu, S. Demarque, P., 2002, ApJ., 576, 963.
\bibitem{} Van den Bergh, S., 1993, ApJ, 411, 178
\bibitem{} VandenBerg, D. A., 2000, ApJS, 129, 315.
\bibitem{} VandenBerg, D. A., Swenson, F. J., Rogers, F. K., Iglesias, C. A., Alexander, D. R., 2000, ApJ, 532, 430.
\bibitem{} van den Marel, R.P., Alves, D.R., Hardy, E., Suntzeff, N.B., 2002, AJ, 124, 2639
\bibitem{} Wallerstein, G., Cox, A.N., 1984, PASP, 96, 677.
\bibitem{} Zinn R., 1985, ApJ, 293, 424.
\bibitem{} Zinn R., 1993, in ASP Conf. Ser. 46, The Golbular Cluster-Galaxy connection. Eds. Graeme H. Smith \& Jean P. Brodie. (San Francisco; ASP) p. 38
\bibitem{} Zinn R., 1996, in ASP Conf. Ser. 92, 
Formation of the Galactic Halo. Inside and Out. Eds. Heather Morrison \& Ata Sarajedini. (San Francisco; ASP), p. 211
\bibitem{} Zinn R., West, M.J., 1984, ApJS, 55, 45.
\bibitem{} Zoccali, M., Renzini, A., Ortolani, S., Greggio, L., Saviane, I., Cassisi, S., Rejkuba, M., Barbuy, B., Rich, R. M., Bica, E., 2003, A\&A, 399, 931.
\end{thebibliography}
\end{document}